\documentclass[preprint,nofootinbib]{revtex4}
\usepackage{epsfig}

\begin{document}
\begin{titlepage}

\begin{center}
{\Large \bf From de Sitter to de Sitter:  A New Cosmic \vskip 0.1cm
Scenario without Dark Energy}

\vskip 0.1cm

J. A. S. Lima$^{a}$ and S. Basilakos$^{b,c}$
\end{center}

\begin{quote}
$^a$Departamento de Astronomia, Universidade de S\~ao Paulo, Rua do
Mat\~ao 1226,
05508-900, S\~ao Paulo, SP, Brazil, e-mail: limajas@astro.iag.usp.br\\
$^b$Academy of Athens, Research Center for Astronomy and Applied
Mathematics, Soranou Efesiou 4, 11527, Athens, Greece, e-mail:
svasil@Academyofathens.gr
$^c$ High Energy Physics Group, Dept. ECM, Univ. de Barcelona,
Av. Diagonal 647, E-08028 Barcelona, Spain
\end{quote}

\centerline{\bf ABSTRACT}
\bigskip

In the present lore of cosmology, matter and space-time emerged from
a singularity and evolved through four different regimes: inflation,
radiation, dark matter and dark energy dominated eras. In the
radiation and dark matter dominated stages, the expansion of the
Universe decelerates while the inflation and dark energy eras are
accelerating regimes. So far there is no clear cut connection
between these accelerating periods. More intriguing, the substance
driving the present accelerating stage remains a mystery, and the
best available candidate ($\Lambda$-vacuum) is plagued with the
coincidence and cosmological constant problems.

In this paper we overcome such problems through an alternative
cosmic scenario based on gravitationally-induced particle
production. The model proposed here is non-singular with the
space-time emerging from a pure initial de Sitter stage thereby
providing a natural solution to the horizon problem. Subsequently,
due to an instability provoked by the production of massless
particles, the Universe evolves smoothly to the standard radiation
dominated era thereby ending the production of radiation as required
by the conformal invariance (Parker's theorem). Next, the radiation
becomes subdominant with the Universe entering in the cold dark
matter dominated era. Finally,  the negative pressure associated
with the creation of cold dark matter particles accelerates the
expansion and drives the Universe to a final de Sitter stage. The
late time cosmic expansion history is exactly like in the standard
$\Lambda$CDM model, however, there is no dark energy. The model
evolves between two limiting (early and late time) de Sitter
regimes. Our scenario is fully determined by two extreme energy
densities, or equivalently, the associated de Sitter Hubble scales
connected by $\rho_I/\rho_f=(H_I/H_f)^{2} \sim 10^{122}$.


\end{titlepage}
\pagestyle{plain} 

The microscopic description for gravitationally-induced particle
production in an expanding Universe began with Schr\"odinger's
\cite{Scho39} seminal paper, who referred to it as an alarming
phenomenon. In the late 1960s, this issue was rediscussed by Parker
and others \cite{Parker,BDbook} based on the Bogoliubov mode-mixing
technique in the context of quantum field theory in curved
space-time. Physically, one may think that the (classical) time
varying gravitational field works like a `pump' supplying energy to
the quantum fields.

In order to understand the basic approach,  let us now consider a
real minimally coupled massive scalar field $\phi$ evolving in a
flat expanding Friedman-Robertson-Walker (FRW) geometry.  The field
is described by the following action\footnote{We adopt units such
that $\hbar=k_B=c=1$.}

\begin{equation}\label{m63}
S={1\over 2} \int \sqrt{-g}d^4 x
\bigg[g^{\alpha\beta}\partial_\alpha\phi\partial_\beta\phi-m^2
\phi^2\bigg]\,.
\end{equation}
In terms of the conformal time $\eta$ ($dt = a(\eta)d\eta$), the
metric tensor $g_{\mu\nu}$ is conformally equivalent to the
Minkowski metric $\eta_{\mu\nu}$, so that the line element is
$ds^2=a^2(\eta)\eta_{\mu\nu}dx^\mu dx^\nu$, where $a(\eta)$ is the
cosmological scale factor. Writing the field $\phi (\eta,x) =
a(\eta)^{-1}\chi$, one obtains from the above action
\begin{equation}\label{m67}
\chi''- \nabla^2 \chi +\bigg( m^2a^2-{a''\over a}\bigg)\chi=0\,,
\end{equation}
where the prime denotes derivatives with respect to $\eta$. Notice
that the field $\chi$ obeys the same equation of motion as a massive
scalar field in Minkowski space-time, but now with a time dependent
{\em effective mass},
\begin{equation}\label{m68}
m^2_{eff}(\eta)\equiv m^2a^2-{a''\over a}\,.
\end{equation}
This time varying mass accounts for the interaction between the
scalar and the gravitational fields.  The energy  of the field
$\chi$ is not conserved (its action is explicitly time-dependent),
and, more important, its quantization leads to particle creation at
the expense of the classical gravitational background
\cite{Parker,BDbook,Muka}.

On the other hand,  in the framework of general relativity theory
(TRG), the scale factor of a FRW type Universe dominated by
radiation satisfies the following equation:

\begin{equation}
a{\ddot a} + {\dot a}^{2} = 0
\end{equation}
or, in the conformal time, $a'' = 0$. Therefore, for  massless
fields ($m=0$),  there is no particle production  since Eq.
(\ref{m67}) reduces to the same of a massless field in Minkowski
spacetime, and, as such,  its quantization becomes trivial. This is the basis of
Parker theorem concerning the absence of massless particle
production in the early stages of the Universe. Note that Parker's
result does not forbid the production of massless particles in a
very early de Sitter stage ($a''\neq 0$). Potentially, we also see  that massive particles
can always be produced by a time varying gravitational field. As we shall see, such features are 
incorporated in the scenario proposed here. 

In principle, for applications in cosmology, the above semiclassical
results has three basic difficulties, namely:

(i) The scalar field was treated as a test field, and, therefore,
the FRW background is not modified by the newly produced particles.

(ii) The particle production is an irreversible process, and, as
such, it should be constrained by the second law of thermodynamics.

(iii) There is no a clear prescription of how an irreversible
mechanism of quantum origin can be incorporated in the Einstein
Field Equations (EFE).

Later on, a possible macroscopic solution for these problems was put
forward by Prigogine and coworkers \cite{Prigogine} using
non-equilibrium thermodynamics for open systems, and by  Calv\~ao,
Lima \& Waga \cite{LCW}  through a covariant relativistic treatment
for imperfect fluids (see also \cite{LG92}). The leitmotiv of the
approach is that particle production, at the expense of the
gravitational field, is an irreversible process constrained by the
usual requirements of non-equilibrium thermodynamics.  This
irreversible process is described by a negative pressure term in the
stress tensor whose form is constrained by the second law of
thermodynamics\footnote{The semiclassical approach is unable to
provide the entropy burst accompanying the particle production since
it is adiabatic and reversible.}.

In comparison to the standard equilibrium equations, the
irreversible creation process is described by two new ingredients: a
balance equation for the particle number density and a negative
pressure term in the stress tensor. Such quantities are related to
each other in a very definite way by the second law of
thermodynamics. Since the middle of the nineties, several
interesting features of cosmologies with creation of cold dark
matter  and radiation have been investigated by many authors
\cite{ZP2,ZP3,ZP4,ZP5}.

In this context, we are proposing here a new cosmological scenario
where the accelerating stages of the cosmic evolution are powered
uniquely by the creation of massless and massive cold dark matter
particles. In this model, the Universe starts from a  de
Sitter dominated phase ($a \propto e^{H_I t}$) powered by the production of massless
particles.  Subsequently, it deflates and evolves to the standard
radiation phase ($a \propto t^{1/2}$) thereby ending  the creation of massless particles.
Due to expansion,  the radiation becomes subdominant with the
Universe entering in the cold dark matter (CDM) dominated era.
Finally,  the negative pressure associated with the creation of cold
dark matter particles accelerates the expansion and drives the
Universe to a final de Sitter stage. The horizon problem is
naturally solved in the initial de Sitter phase. In addition, the
transition from Einstein-de Sitter ($a \propto t^{2/3}$) to a de Sitter final stage ($a \propto e^{H_f t}$) 
guarantee the consistence of the model with the supernovae type Ia
data and complementary observations.  A transition redshift of the
order of a few (exactly the same value predicted by $\Lambda$CDM) is
also obtained.

For simplicity, let us consider the EFE for a flat geometry:
\begin{equation}
\label{fried}
    8\pi G\rho = 3 \frac{\dot{a}^2}{a^2},
\;\;\;\;\;\;   8\pi G(p + p_{c})=  -2 \frac{\ddot{a}}{a} -
\frac{\dot{a}^2}{a^2},
\end{equation}
where an overdot means time derivative, $\rho$ and p are the
dominant energy density and pressure of the cosmic fluid,
respectively, and $p_c$ is a dynamic pressure which depends on the
particle production rate. Special attention has been paid to the
simpler process termed  ``adiabatic" particle production. It means
that particles and entropy are produced in the space-time, but the
specific entropy (per particle), $\sigma = S/N$, remains constant
\cite{LCW}. In this case,  the creation pressure reads [6-9]
\begin{equation}
\label{CP}
    p_{c} = -\frac{(\rho + p) \Gamma}{3H},
\end{equation}
where $\Gamma$ with dimensions of $(time)^{-1}$ is the particle
production rate and $H={\dot a}/a$ is the Hubble parameter.


{\it How the evolution of $a(t)$ is affected by $\Gamma$?} By
assuming a dominant cosmic fluid satisfying the equation of state
(EoS), $p=\omega\rho$, where $\omega$ is a constant, the EFE imply
that
\begin{equation}
\label{varH0} \dot H + \frac{3}{2} (1 + \omega)  H^2
\left(1-\frac{\Gamma}{3H}\right) = 0.
\end{equation}
The de Sitter solution ($\dot H=0$,  $\Gamma = 3H = constant$) is
now possible regardless of the  EoS defining the cosmic fluid. Since
the Universe is evolving, such a solution is unstable, and, as long
as $\Gamma << 3H$, conventional solutions without particle
production are recovered.

$\Rightarrow$ {\it The main effect of $\Gamma$ is to provoke a
dynamic instability in the space-time, thereby allowing a transition
from de Sitter to a conventional solution, and vice versa.}

{\bf A. From an early de Sitter stage to the standard radiation
phase}

\vskip 0.1cm

Let us first discuss the transition from an initial de Sitter stage
to the standard radiation phase. The main theoretical constraints
are:

\begin{itemize}

 \item{\it The model must not only solve the horizon problem  but also provide a
 quasiclassical boundary condition to quantum cosmology (a hint on how to solve the initial singularity problem).}

\item{\it Massless particles  cannot be quantum-mechanically produced in the conventional radiation phase (Parker's Theorem).}

\end{itemize}
To begin with, let us  assume a radiation dominated Universe
($\omega=1/3,\, \Gamma\equiv\Gamma_r $). The dynamics is determined
by the ratio $\Gamma_r/3H$ (see (\ref{varH0})).  The most natural
choice would be a ratio which favors no epoch in the evolution of
the Universe ($\Gamma_r/3H = constant$). However, the particle
production must be strongly suppressed, $\Gamma_r/3H << 1$, when the
Universe enters the radiation phase. This means that the expansion The simplest formula satisfying
such a criterion is linear, namely: $\Gamma_r/3H = H/H_I$, where $H_I$  is the initial de Sitter expansion rate ($H \leq H_I$).
Inserting this into (\ref{varH0})  it becomes:

\begin{equation}
\label{varH1} \dot H + {2}  H^2 \left(1-\frac{H}{H_I}\right) = 0.
\end{equation}

The solution of the above equation  can be written as

\begin{equation}
\label{solH} H(a) = \frac{H_I}{1+ D a^{{2}}},
\end{equation}
where $D \geq 0$ is an integration constant. Note that $H=H_I$ is a
special solution of Eq.(\ref{varH1}) describing the exponentially
expanding de Sitter space-time. This solution is  unstable with
respect to the critical value $D=0$. For $D>0$, the universe starts
without a singularity and evolves continuously towards a radiation
stage, $a \sim {t}^{1/2}$, when $Da^{2} >> 1$. By integrating
(\ref{solH}), we obtain the scale factor:

\begin{equation}
H_I t=\ln \frac{a}{a_*}+\frac{\lambda^{2}}{2}{(a/a_{*})}^{2},
\label{full1}
\end{equation}
where $\lambda^{2}=Da_*^{2}$ is an integration constant and $a_*$
defines the transition from the de Sitter stage to the beginning of
the standard radiation epoch\footnote{This kind of evolution was
first discussed by G. L. Murphy \cite{Murphy73} by studying possible
effects of the second viscosity in the very early Universe. Later
on, it was also investigated in a more general framework involving
cosmic strings by J. D. Barrow \cite{Barrow88} who coined the
expression ``Deflationary Universes".  It has also been discussed in
connection with decaying $\Lambda(t)$-models \cite{LT96}.}. At early
times ($a \ll a_{*}$), when the logarithmic term dominates, one
finds $ a \simeq  a_*e^{H_I t}, $ while at late times, $a \gg a_*, H
\ll H_I$, (\ref{full1}) reduces to $a \simeq a_*
\left(\frac{2H_I}{\lambda^{2}}t\right)^{1/2}$, and the  standard radiation phase is reached.

It should be noticed that the time scale ${H_I}^{-1}$ provides the greatest value of the
energy density, $\rho_{I}=\frac{3H_{I}^{2}}{8{\pi}G}$,
characterizing the initial de Sitter stage which is supported by the
maximal radiation production rate, $\Gamma_r=3H_{I}$.  From
(\ref{fried}) and (\ref{solH}) we obtain the radiation energy
density:

\begin{equation}\label{rho1}
\rho_r={\rho_{I}}\left[1 +
\lambda^{2}\left(\frac{a}{a_*}\right)^{2}\right]^{-2}.
\end{equation}
As expected, we see again that the conventional radiation phase, $\rho_r \sim a^{-4}$,
is attained when $a \gg a_*$.

{\it How the cosmic temperature evolves?} For ``adiabatic" particle
production the energy density scales as $\rho_{r} \sim T^{4}$
\cite{LCW,LG92}, and the above equation implies that

\begin{equation}\label{T}
T_r={T_{I}}\left[1 +
\lambda^{2}\left(\frac{a}{a_*}\right)^{2}\right]^{-1/2},
\end{equation}
where $T_I$ is the temperature of the initial de Sitter phase which
must be uniquely determined by the scale $H_I$. We see that the
expansion proceeds isothermally during the de Sitter phase
($a<<a_*$). A basic consequences is:

$\Rightarrow$ {\it The supercooling and subsequent reheating taking
place in several inflationary variants are avoided \cite{Turner83,KT90}. In other words, there is no the so-called 
`graceful exit' problem.}

After de Sitter stage, the temperature decreases continuously in the
course of the expansion. For $a>>a_*$ ($H<<H_I$), we obtain $T \sim
a^{-1}$. Accordingly, the comoving number of photons becomes
constant since $n \propto {a}^{-3}$, as expected for the standard
radiation stage.\footnote{Since $n_r \propto T^{3}$,  the average
photon concentration reads $n_r=n_{I}\left[1 +
\lambda^{2}\left(\frac{a}{a_*}\right)^{2}\right]^{-\frac{3}{2}}$.}

{\it What about the initial temperature $T_I$?} Since the model
starts as a de Sitter space-time, the most natural choice is to
define $T_I$ as the Gibbons-Hawking temperature \cite{GibHaw} of its
event horizon, $T_I=H_{I}/2\pi$.
Naively, one may expect $T_I$ of the same order or smaller than the
Planck temperature because of the classical description. From EFE we
have $\rho_I = 3{m_{Pl}}^{2}{H_I}^{2}/8 \pi$ (where $m_{Pl}\simeq
1.22\times 10^{19} GeV$), and since the energy density is $\rho_I =
N_*(T)T_I^{4}$, one finds  $T_I \sim H_I \sim 10^{19} GeV$ (where
$N_* (T) = {\pi}^2 g_*(T)/30$ depends on the number of effectively
massless particles).

Naturally, due to the initial de Sitter phase, the model is free of particle horizons. A light pulse beginning at
$t=-\infty$ will have traveled by the cosmic time $t$ a physical
distance, $d_{H}(t)=
a(t)\int_{-\infty}^{t}\frac{d\tilde{t}}{a(\tilde{t})}$, which
diverges thereby implying the absence of particle horizons:

$\Rightarrow$ {\it The local interactions may homogenize the whole
Universe.}




Since photons are not produced in the radiation phase, the Big-Bang Nucleosynthesis (BBN) may work in
the conventional way \cite{Steigman}.  Subsequently, the Universe enters the cold
dark matter (Einstein-de Sitter, $a(t)\propto t^{2/3}$) dominated
phase.

{\bf B. From Einstein-de Sitter to a late time de Sitter stage}

Due to conservation of baryon number the remaining question is the
production rate of cold dark matter particles and the overall late
time evolution.  In other words, what is the form of $\Gamma_{dm}$?
For simplicity, we consider here only the dominant CDM component.

In principle, $\Gamma_{dm}$ should be determined  from quantum field
theory in curved spacetimes.  In the absence of a rigorous
treatment, we consider  (phenomenologically)  the  following fact
\cite{K08,Peebles,Kom11}:

\begin{itemize}
\item{\it All available observations  are in accordance with the $\Lambda$CDM evolution both at the
background and perturbative levels.}
\end{itemize}
Now, we recall that a flat $\Lambda$CDM model evolves like:
\begin{equation} \label{LCDM}
\dot H + \frac{3}{2} H^2 \left[1 -
\left(\frac{H_f}{H}\right)^{2}\right]=0,
\end{equation}
where ${H_f}^{2} = \Lambda/3$ sets the Hubble scale of the final de
Sitter stage ($H \geq H_f$). Such  behavior should be compared to
that predicted for a dust filled model ($\omega = 0,\,
\Gamma\equiv\Gamma_{dm}$) with particle production (see Eq.
(\ref{varH0})):
\begin{equation}
\label{varCDM} \dot H + \frac{3}{2} H^2 \left(1-
\frac{\Gamma_{dm}}{3H} \right) = 0.
\end{equation}
By comparing (\ref{LCDM}) and (\ref{varCDM}),  we see that the same
background evolution requires that $\Gamma_{dm}/{3H}=
\left({H_f}/{H}\right)^{2}$.
The limiting value of the creation rate, $\Gamma_{dm}=3H_f$, leads
to a late time de Sitter phase ($\dot H=0,\, H=H_f$) thereby showing
that the de Sitter solution now becomes  an attractor at late times.
With this proviso, the solution of (\ref{varCDM}) reads:

\begin{figure*}
\centerline{\epsfig{file=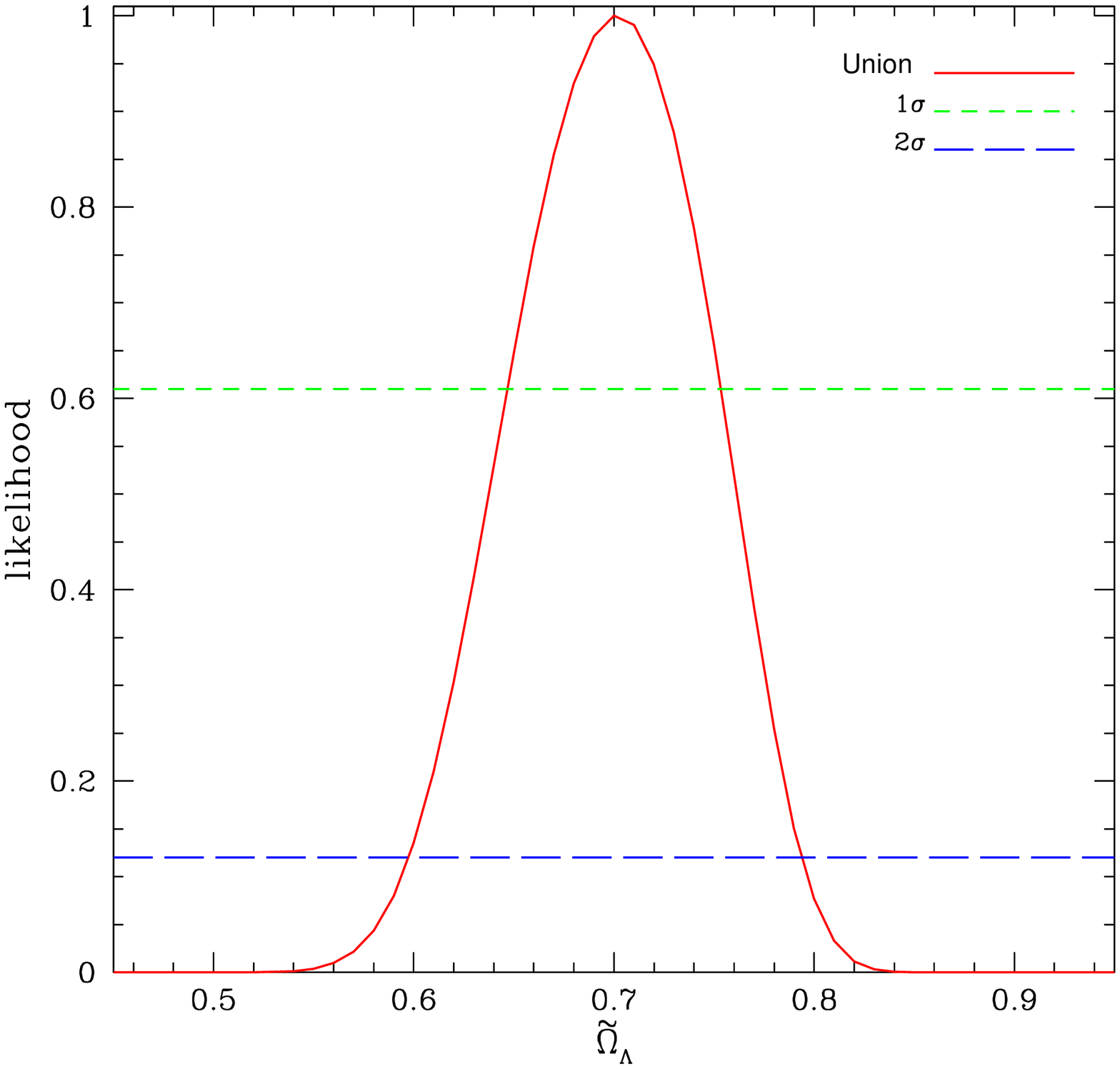,width=78mm, height=56mm}
\epsfig{figure=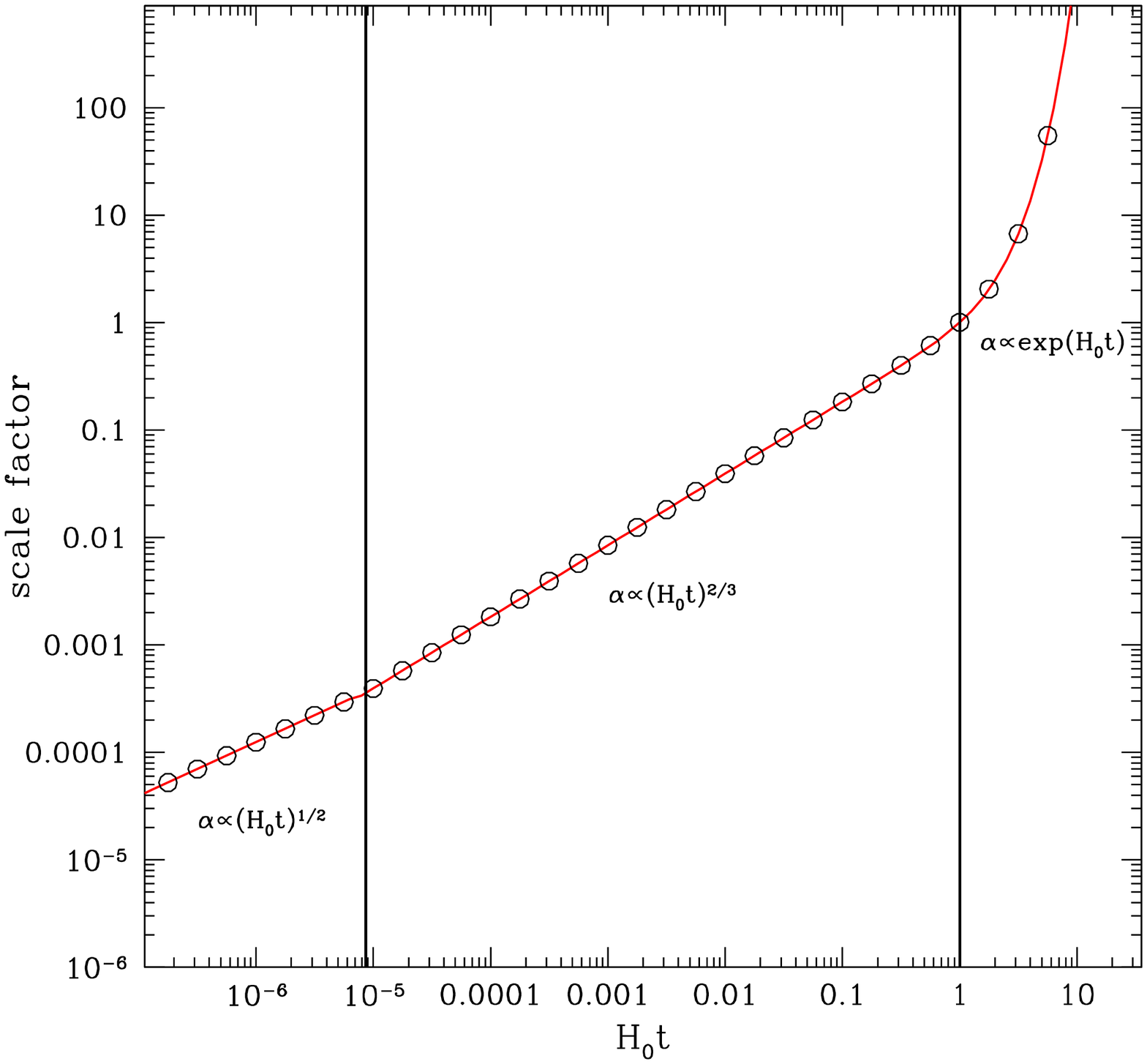, width=78mm, height=56mm} \hskip 0.1in}
\caption{{\bf (a)} The likelihood for $\tilde{\Omega}_{\Lambda}$
based on 307 Supernova data (Union)\cite{K08}. {\bf (b)} Evolution
of the scale factor predicted by the matter creation model (solid
line) and the traditional $\Lambda$CDM cosmology (open points).  In
this plot we have adopted the best fit,
$\tilde{\Omega}_{\Lambda}=0.72$, from Supernova data.} \label{fig1}
\end{figure*}



\begin{equation}
\label{Hz}
 {H}^{2} = {H_0}^2 \left[(1-\tilde{\Omega}_{\Lambda})(1+z)^3+\tilde{\Omega}_{\Lambda} \right],
\end{equation}
where $\tilde{\Omega}_{\Lambda} \equiv (H_f/H_0)^{2}$ is  smaller
than unity and $1+z=a^{-1}$. Such solution mimics the Hubble
function $H(z)$ of the traditional flat $\Lambda$-cosmology, with
$\tilde{\Omega}_{\Lambda}$ playing the dynamical role of
$\Omega_{\Lambda}$ (dark energy appearing in the concordance model).
The dark matter parameter ($\Omega_{dm}=1$) is also replaced by an
effective parameter, $({\Omega_{dm}})_{eff} \equiv
1-\tilde{\Omega}_{\Lambda}$, which quantifies the amount of matter
that is  clustering. This explains why this model is in agreement
with the dynamical determinations related to the amount of the cold
dark matter at the  cluster scale, and, simultaneously, may also be
compatible with the position of the first acoustic peak in the
pattern of CMB anisotropies which requires $\Omega_{total}=1$.

By integrating  (\ref{Hz}) we obtain:

\begin{equation}
a(t)=\left(\frac{1-\tilde{\Omega}_{\Lambda}}{\tilde{\Omega}_{\Lambda}}\right)^{1/3}
\sinh^{\frac{2}{3}}\left(\frac{3H_{0}\sqrt{\tilde{\Omega}_{\Lambda}}
}{2}t\right).
\end{equation}

$\Rightarrow$ {\it The late time dynamics is determined by a single
parameter ($\tilde{\Omega}_{\Lambda}$) and is identical to that
predicted by the flat $\Lambda$CDM model.}

In Figure 1, we show the likelihood of $\tilde{\Omega}_{\Lambda}$
based on the Union supernova sample. Note also that by replacing
the value of $\Gamma_{dm}$ into the definition of the creation
pressure (see Eq. \ref{CP}) one obtains that it is negative and
constant ($p_c = - 3H_f^{2}/ 8\pi G =
-3\tilde{\Omega}_{\Lambda}H_0^{2}/8\pi G$). Therefore, the late time
evolution of our complete cosmological scenario coincides exactly
with the one recently discussed in Refs. \cite{LJO,BL10} following a slightly
different approach.

Concluding, a new cosmology based on the production of massless
particles (in the early de Sitter phase) and CDM particles (in the
transition to a late time de Sitter stage) has been discussed. The
same mechanism avoids the initial singularity, particle horizon and
the late time coincidence problem of the $\Lambda$CDM model has been
eliminated ($\Lambda \equiv 0$).  

In this scenario, the standard cosmic phases - a  radiation era
followed by an Einstein-de Sitter evolution driven by
nonrelativistic matter until redshifts of the order of a few - are
not modified. However, the model has two extreme accelerating phases
(very early and late time de Sitter phases) powered by the same
mechanism (particle creation). Therefore, it sheds some light on a
possible connection among the different accelerating stages of the
universe. In particular, since $H_f^{2} = \tilde{\Omega}_{\Lambda}
H_0^{2}$, where $\tilde{\Omega}_{\Lambda}\sim 0.7$ and $H_{0}\simeq
1.5\times 10^{-42}GeV$, it sets the ratio of the primeval and late
time de Sitter scales to be $\rho_{I}/\rho_{f}=(H_I/H_f)^{2} \sim
10^{122}$. Such a result in the present context has no correlation
with the so-called cosmological constant problem
\cite{LT96,weinberg}.

As it appears, the cosmic history discussed here is semi-classically
complete. However, there is no guarantee that the initial de Sitter
configuration is not only the boundary condition of a true quantum
gravitational effect. In other words, the very early de Sitter phase
may be the result of a quantum fluctuation which is further
semi-classically  supported by the creation of massless particles
(in this connection see \cite{Pri2} and Refs. there in).

Naturally, the existence of an early isothermal de Sitter phase
suggests that thermal fluctuations (within the de Sitter event
horizon) may be the causal origin of the primeval seeds that will
form the galaxies. Such a possibility  and its consequences for the
structure formation problem deserves a closer investigation and is
clearly out of the scope of the present paper.

 At present, we also
know that a more complete version of the late time evolution must be
filled with CDM ($\sim$ 96\%) and baryons ($\sim$ 4\%), and, unlike
$\Lambda$-cosmology, the baryon to dark matter ratio is a redshift
function \cite{LJO,BL10}. In particular, this means that studies
involving  the gas mass fraction may provide a crucial test of our
scenario, potentially, modifying our present view of the dark
sector. Some investigations along the above discussed lines are in
progress and will be published elsewhere.

\vspace{0.3cm} {\bf Acknowledgments:} The authors are grateful to G.
Steigman, A. C. C. Guimar\~aes, J. F. Jesus and  F. O. Oliveira  for
helpful discussions. JASL is partially supported by CNPq and FAPESP under grants 304792/2003-9 and 04/13668-0,
respectively. SB wishes to thank the Dept. ECM of the 
University of Barcelona for the hospitality, and the financial support from the
Spanish Ministry of Education, within the program of Estancias de
Profesores e Investigadores Extranjeros en Centros Espanoles (SAB2010-0118).

\end{document}